\documentstyle[emulateapj,psfig]{article}

\newcommand{\pcc}{\,{\rm cm}^{-3}}

\newcommand{\etal}{et~al.\ }
\def\kms{\,{\rm km\,s^{-1}}}
\def\kmsmpc{\,{\rm km\,s^{-1}\,Mpc^{-1}}}
\def\msun{\,{\rm M_\odot}}
\def\rsun{\,{\rm R_\odot}}
\def\mden{\,{\rm M_\odot\,Mpc^{-3}}}

\def\ergss{\,{\rm erg\,s^{-1}}}
\def\spose#1{\hbox to 0pt{#1\hss}}
\def\lta{\mathrel{\spose{\lower 3pt\hbox{$\mathchar"218$}} \raise 2.0pt\hbox{$\mathchar"13C$}}}
\def\gta{\mathrel{\spose{\lower 3pt\hbox{$\mathchar"218$}} \raise 2.0pt\hbox{$\mathchar"13E$}}}

\def\ni{\noindent}

\def\vir{{\rm vir}}

\newcommand\beq{\begin{equation}}
\newcommand\eeq{\end{equation}}

\lefthead{Madau \& Rees}
\righthead{Pregalactic Black Holes}
\submitted{ApJ, in press}
\makeatletter

\newenvironment{figurehere}
  {\def\@captype{figure}}
  {}
\makeatother

\begin{document}

\title{Massive Black Holes as Population III Remnants}
\author{Piero Madau\altaffilmark{1,2} and Martin J. Rees\altaffilmark{3}}

\altaffiltext{1}{Department of Astronomy and Astrophysics, University of
California, Santa Cruz, CA 95064.}
\altaffiltext{2}{Osservatorio Astrofisico di Arcetri, Largo Enrico Fermi 5,
50125 Firenze, Italy.}
\altaffiltext{3}{Institute of Astronomy, Madingley Road, Cambridge CB3 0HA, UK.}

\begin{abstract}
Recent numerical simulations of the fragmentation of primordial molecular
clouds in hierarchical cosmogonies have suggested that the very first 
stars (the so--called Population III) may have been rather massive.  
Here we point out that a numerous population of massive black holes (MBHs) --
with masses intermediate between those of stellar and supermassive holes -- 
may be the endproduct of such an episode of pregalactic star formation.
If only one MBH with $m_\bullet\gta 150\,\msun$ formed in each of the `minihalos'
collapsing at $z\approx 20$ from 3--$\sigma$ fluctuations, then the mass density
of Pop III MBHs would be comparable to that of the supermassive variety 
observed in the nuclei of galaxies.
Since they form in high--$\sigma$ rare density peaks, relic MBHs are 
predicted to cluster in the bulges of 
present--day galaxies as they become incorporated through a series of 
mergers into larger and larger systems. 
Dynamical friction would cause $\gta 50\,(m_\bullet/150\,\msun)^{1/2}$ such 
objects to sink towards the 
center. The presence of a small cluster of MBHs in galaxy nuclei may 
have several interesting consequences associated with tidal captures of ordinary 
stars (likely followed by disruption), MBH capture by the central supermassive black 
hole, gravitational wave radiation from such coalescences.
Accreting pregalactic MBHs may be detectable as ultra--luminous, off--nuclear 
X--ray sources.

\end{abstract}
 
\section{Introduction}
The very first generation of stars must have formed
out of unmagnetized, metal--free gas, characteristics leading to a 
well--defined set of cosmic initial conditions, a significant simplification of the
relevant physics, and an initial stellar mass function (IMF) which may have 
been very different from the present--day Galactic case (Larson 1998).
The primordial IMF is expected to play a crucial role in determining both 
the impact of Pop III stars on the ionization, thermal, and chemical 
enrichment history of the intergalactic medium (Couchman \& Rees 1986), and 
the direct observability of the universe at the end of the `dark ages', e.g.,
through transient sources such as supernovae or $\gamma$--ray bursts 
(Barkana \& Loeb 2001; and references therein). 
In the absence of metals, it is molecular hydrogen (H$_2$) which acts as the 
main gas coolant in some of the earliest non--linear 
cosmic structures expected in hierarchical clustering theories, subgalactic 
fragments (`minihalos') with total masses $M\gta$ a few$\times 10^5\msun$ 
collapsing from high--$\sigma$ fluctuations at redshift $z\gta 20$.
Recent three--dimensional numerical simulations of the fragmentation of primordial 
molecular clouds in these popular cosmological scenarios -- all variants of the 
cold dark matter (CDM) cosmogony -- have shown convergence 
towards a thermal regime where the gas density is $\sim 10^4\,\pcc$ and the 
temperature is a few hundred Kelvin (Bromm, Coppi, \& Larson 1999, 2001; 
Abel, Bryan, \& Norman 2000). Below 
this temperature the ro--vibrational line cooling time by H$_2$ is longer 
than the collapse time, while above this density
the excited states of hydrogen molecules are in local thermodynamic 
equilibrium and the cooling time is nearly independent of density. At this 
point collapse becomes quasistatic rather than a runaway free--fall, and all 
simulations show the formation of one or several Jeans unstable clumps with 
masses exceeding a few hundred solar masses. 
Because of the slow subsonic contraction, the gas distribution is found to
remain smooth prior to the onset of three body H$_2$ formation (densities in 
excess of
$10^8\,\pcc$), and no further fragmentation into sub--components is seen.
This suggests that the very first stars may have 
been very massive (Bromm \etal 1999; Abel \etal 2000), as or even more 
massive than the Pistol Star, $m\gta 200\,\msun$ (Figer \etal 1998).

In practice, the characteristic mass of the first stars and the fraction of the 
halo gas which is available for early star formation will depend on 
radiative line transfer effects which are difficult to simulate in 
multi--dimensions, the poorly understood physics of stellar radiative and 
mechanical `feedback' mechanisms, and perhaps also on the chemistry of 
HD and LiH molecules (Galli \& Palla 1998). 
If each clump in the above simulations ends up as a single 
very massive star (VMS) with $m\gta 250\,\msun$, then a numerous population of 
primordial black holes with masses $m_\bullet\approx m/2\gta 150\,\msun$ 
is the expected endproduct of such an episode of pregalactic star formation 
(e.g., Bond, Arnett, \& Carr 1984; Larson 1999; Schneider \etal 2000; 
Fryer, Woosley, \& Heger 2000). The early work by Bond \etal (1984) was 
motivated by the possibility that the remnants of VMSs could be an important 
contributor to the dark matter in galaxies (see also Carr 1994). The holes 
envisaged here, in contrast,
collectively contribute a very small fraction of the nucleosynthetic baryons.   
Since they form in high--$\sigma$ peaks, these relic, massive black holes 
(MBHs) are predicted to be strongly `biased', i.e. more abundant in 
the central parts of present--day galaxies. The properties and detectability of 
primordial MBHs are the main focus of this {\it Letter}.

\section{The fate of VMS\lowercase{s}}
The evolution and stability of zero--metallicity VMSs ($Z=0; m=120$--300$\,
\msun$) have been recently investigated by Fryer \etal (2000) and Baraffe, 
Heger, \& Woosley (2001). Above $100\,\msun$, main sequence stars of 
Pop I composition are known to be vibrationally unstable to nuclear--powered 
radial pulsations, an instability which leads to substantial mass loss. In 
$Z=0$ VMSs, however, the efficiency of these instabilities appears to be 
greatly reduced relative to their metal--rich counterparts due to their hotter
cores. Since, at zero metallicity, 
mass loss through radiatively--driven stellar winds is also expected to be 
negligible (Kudritzki 2000), Pop III stars may actually die losing only 
a small fraction of their mass. If they retain their large mass until death, VMSs 
with $100\lta m\lta 250\,\msun$ (helium cores below about $150\,\msun$) will 
encounter the 
electron--positron pair instability and disappear in a giant nuclear--powered
explosion, leaving no compact remnants. In still heavier stars, however, oxygen and 
silicon burning is unable to drive an explosion, and complete collapse 
to a black hole will occur (Bond \etal 1984). 

\section{Formation of pregalactic MBH\lowercase{s}}
While considerable uncertainties still remain on the characteristic mass or mass
spectrum of the first luminous objects, there is growing theoretical basis for 
considering
the possibility that a population of metal--free stars with masses larger 
than $250\,\msun$ formed in the minihalos that populated the universe 
at high redshifts. After 2 Myr such a star would have collapsed to a MBH 
containing at least half of the initial stellar mass (Fryer \etal 2000).

In a CDM flat cosmology with $\Omega_M=1$, $H_0=100\,h\,\kmsmpc$, and
rms mass fluctuation normalized at present to $\sigma_8=0.63$ on spheres of 
$8\,h^{-1}\,$Mpc, minihalos of mass $M=3\times 10^5\,h^{-1}\,\msun$\footnote{This 
is well above the cosmological (gas$+$dark matter) Jeans mass for an Einstein--de 
Sitter universe, $M_J=$ $10^{3.8}$ $h^{-1}$ $\msun$ $(\Omega_bh^2/0.02)^{-3/5}$ 
$[(1+z)/20]^{3/2}$. It corresponds instead to the minimum mass 
threshold for baryonic condensation: above this mass the H$_2$ cooling time is
shorter than the Hubble time at virialization, the baryonic material in the
central regions becomes self--gravitating, and stars can form. Recent numerical
simulations
by Fuller \& Couchman (2000) find a minimum collapse mass of $\approx 5\times 
10^5\,\msun$ ($h=0.65$) at $z=20$.}~would be collapsing at 
$z=22$ from 3--$\sigma$ fluctuations. A halo of mass $M$ at this epoch is 
characterized by a virial radius (i.e. the radius of the sphere encompassing a 
mean overdensity of 200) $r_\vir=60\,{\rm pc}~M_{5.5}^{1/3}\,h^{-1}$,
and a circular velocity $V_c=5\,{\kms}~M_{5.5}^{1/3}$ at $r_\vir$.
The gas collapsing along with the dark matter perturbation will be shock heated 
to the virial temperature $T_\vir=1500\,{\rm K}~M_{5.5}^{2/3}\,\mu$,
where $M_{5.5}$ is the halo mass in units of $3\times 10^5 h^{-1}\,\msun$ and $\mu$ 
is the mean molecular weight. The total 
gas mass fraction within the virial radius is equal initially to 
the baryonic matter density parameter $\Omega_b$, with $\Omega_bh^2=0.02\,$.
Also, in a Gaussian theory, 
halos more massive than the 3--$\sigma$ peaks contain 0.3\% of the mass of 
the universe. Assuming only one black hole of mass $m_\bullet\gta 150\,\msun$ forms 
in each of the 3--$\sigma$ minihalos, one estimates that a fraction 
\beq
f_{\rm mbh}={0.003 \, m_\bullet\over 3\times 10^5\,h^{-1}\,\Omega_b\,\msun}\gta
8\times 10^{-5}\,h^3
\eeq
of the nucleosynthetic baryons would end up in pregalactic MBHs.
Interestingly, this is comparable to the mass fraction of the supermassive 
variety found in the nuclei of most nearby galaxies,
\beq
f_{\rm smbh}={\rho_{\rm smbh}\over \rho_{\rm crit}\Omega_b}={6\times 10^5\,h\,
\mden\over 5.5\times 10^9\,\mden}\approx 10^{-4}\,h,
\eeq
where $\rho_{\rm crit}$ is the critical density today and we have taken 
for $\rho_{\rm smbh}$ the value recently inferred by Merritt \& Ferrarese 
(2001). If an extreme top--heavy IMF is linked, as suggested by numerical 
simulations, to primordial H$_2$ chemistry and cooling, then it seems 
unlikely that the formation of MBHs from zero--metallicity VMSs could be a much 
more efficient process, since: 

\smallskip
(1) Metal--free VMSs
emit about $10^{48}$ photons per second per solar mass of stars in the 
Lyman--Werner (LW, 11.2--13.6 eV) bands (Bromm, Kudritzki, \& Loeb 2000).
Over a main--sequence lifetime of $2\times 10^6\,$yr, $10^5$ LW photons will 
be radiated per stellar baryon, or approximately $(10^5\times f_{\rm mbh}\times 
2)\gta $ a few LW photons per baryon in the universe {\it from the 
3--$\sigma$ peaks alone}. 
Radiation in the LW bands will photo--dissociate H$_2$ elsewhere within the 
host halo via the Solomon process (Omukai \& Nishi 1999), escape into the 
intergalactic medium (IGM), and 
form a soft UV cosmic background which suppresses molecular cooling throughout the
universe (Haiman, Rees, \& Loeb 1997), except perhaps in the cores of already 
collapsing clouds where H$_2$ self--shielding may be important.
This background would inhibit the formation of a much more numerous population of 
VMSs, e.g. in $M=3\times 10^5\,h^{-1}\,\msun$ 
minihalos at $z=14$ when they collapse from 2--$\sigma$ fluctuations 
(containing 5\% of the mass of the universe).

\smallskip
(2) The same VMSs in the 3--$\sigma$ peaks also emit similar amounts of 
Lyman--continuum photons, again about $10^5$ per stellar baryon (this is $\gta 20$ 
times higher than the corresponding number for a stellar population with a standard
Salpeter IMF). Photoionization by an internal 
UV source or a nearby external one will heat the 
gas inside a minihalo to a temperature of $10^4\,$K, leading to the 
photoevaporation of baryons out of their hosts. A few 
H--ionizing photons per nucleosynthetic baryon may not be enough to 
photoevaporate all minihalos and reionize the universe by these early times 
(Haiman, Abel, \& Madau 2001), an event which 
must likely wait for the formation, collapse, and cooling of the more 
numerous 2 and 1--$\sigma$ peaks or of halos with higher mass.
After the reionization epoch further fragmentation and
star formation is only possible within more massive halos with virial 
temperatures $T_\vir\gta$ a few $\times 10^4\,$K, where pressure support is 
reduced and gas can condense due to atomic line cooling. 

\smallskip
(3) Stars with masses in the range $100\lta m\lta 250\,\msun$ are predicted
to make pair--instability supernovae 
with explosion (kinetic) energies of up to $10^{53}\,$erg. This is typically
much larger than the baryon binding energy of a subgalactic fragment, 
$E_b\approx 10^{50}\,{\rm erg}\,h^{-1}\,\Omega_b\,M_{5.5}^{5/3}\,[(1+z)/20]$.
Minihalos will then be completely disrupted (`blown--away') by such energetic 
events, metal--enriched material will be returned to the IGM (e.g. Madau, 
Ferrara, \& Rees 2001) and later collect in the cores
of more massive halos formed by subsequent mergers, where a `second' generation
of stars may now be able to form with an IMF which is less biased towards very
high stellar masses.

\section{Pregalactic MBH\lowercase{s} in galaxy bulges}
We have seen that only a small fraction of the baryons need condense 
into VMSs in order to reheat the IGM and choke off the further formation of 
similar objects. The exact threshold depends on uncertain details, but the 
estimates in the previous section suggest that only the baryons within 
3--$\sigma$ peaks would be involved. 
The expected abundance of MBHs -- assuming they behave similarly to 
collisionless dark matter particles -- is then a function of the 
environment, being higher (positively `biased') within a massive 
galaxy bulge that collapsed at early times, and lower (`antibiased') within
a dwarf galaxy collapsing at the present--epoch (White \& Springel 2000; 
Miralda--Escud\'e 2000). This effect can be quite easily quantified within 
the extended Press--Schechter formalism by computing the average number of 
progenitor 
minihalos at redshift $z$ in a unit range of $\ln M$ which will have 
merged at a later time $z_0$ into a more massive halo of mass $M_0$,
\beq
{dN\over d\ln M}=\sqrt{2\over \pi}~M_0\,\sigma_M\,{D\over S^{3/2}}\,
\exp\left[-{D^2\over 2S}\right]\,{d\sigma_M\over dM}
\eeq
where $D=\delta_c(z-z_0)$ and $S=\sigma_M^2-\sigma_{M_0}^2$ (Bower 1991; 
Lacey \& Cole 1993). Here $\sigma^2_M$ ($\sigma_{M_0}^2$) is the 
variance of the linear power spectrum at $z=0$ smoothed with a `top--hat' 
filter of mass $M$ ($M_0$), and $\delta_c=1.686$ is
the usual critical overdensity for spherical collapse in an Einstein--de Sitter
universe. Denoting with $\nu\equiv \delta_c(1+z)/\sigma_M$ and
$\nu_0\equiv \delta_c(1+z_0)/\sigma_{M_0}$ the `heights' of the final
density field in units of the rms, the halo number density 
can be rewritten as
\beq
{dN\over d\ln M}\propto \exp\left[-\nu^2{(1-\alpha\nu_0/\nu)^2\over 
2(1-\alpha^2)}\right],
\eeq  
where $\alpha\equiv \sigma_{M_0}/\sigma_M<1$. A minihalo will
collapse sooner if it lies in a region of larger--scale overdensity,
$\nu_0\gta \alpha\nu/2$, 
leading to an enhanced abundance of MBHs with respect to the mean.
The bias is expected to be weak, e.g., in a `Milky Way' halo 
($M_0=10^{12}\,h^{-1}\, \msun$, $\sigma_{M_0}=3$) if this collapsed at 
$z_0=0.8$ from a 1--$\sigma$ fluctuation ($\nu_0=1$). In this case the 
abundance of $M=3\times 10^5\,h^{-1}\,\msun$
minihalos formed at $z=22$ from 3--$\sigma$ fluctuations ($\nu=3$), that 
become part of our model Milky Way at $z_0$, would actually correspond to a 
density peak equal to $(\nu-\alpha\nu_0)/(1-\alpha^2)^{1/2}=2.84$ times the 
rms, or an enhancement of only a factor 1.6 relative to the mean. Assume, 
on the other hand, that galaxy bulges form at early times. In a bulge 
of total mass $M_0=2\times 10^{11}\,h^{-1}\,\msun$ ($\sigma_{M_0}=3.9$) 
collapsing at $z_0=3.5$ from a 2--$\sigma$ 
fluctuation, the abundance of minihalos would 
actually correspond to a density peak equal to 2.5 times the rms, or an 
enhancement factor of about 4.  
Figure 1 shows the number of minihalos -- hence of MBHs -- of mass $M$
formed at different redshifts $z$ which end up as part of a 
$M_0$ system at $z_0$, for $(z_0, M_0)=0.8, 10^{12}\,h^{-1}\,\msun$ 
(`Milky Way' halo) and $(z_0, M_0)=3.5, 2\times 10^{11}\,h^{-1}\,\msun$ 
(older `bulge'). 

\section{Pregalactic MBH\lowercase{s} in galaxy nuclei}
We have shown that, if current structure formation theories are correct and 
the primordial IMF was extremely top--heavy, a large number of MBHs -- the 
remnants of pregalactic VMSs born at very high redshifts in rare high--$\sigma$ 
peaks -- is expected to populate the bulges of present--day galaxies. As they 
get incorporated through hierarchical merging into more massive  
halos, these MBHs will be surrounded by dense baryonic (stars$+$gas) cores 
which will be gradually dragged into the center by dynamical friction against 
the visible and dark matter background. Dynamical friction will favor 
accretion of larger mass subunits and infalling satellites with low initial 
angular momenta (high eccentricities): the cores will merge and undergo 
violent relaxation.   

\begin{figurehere}
\vspace{-0.8cm}
\centerline{
\psfig{figure=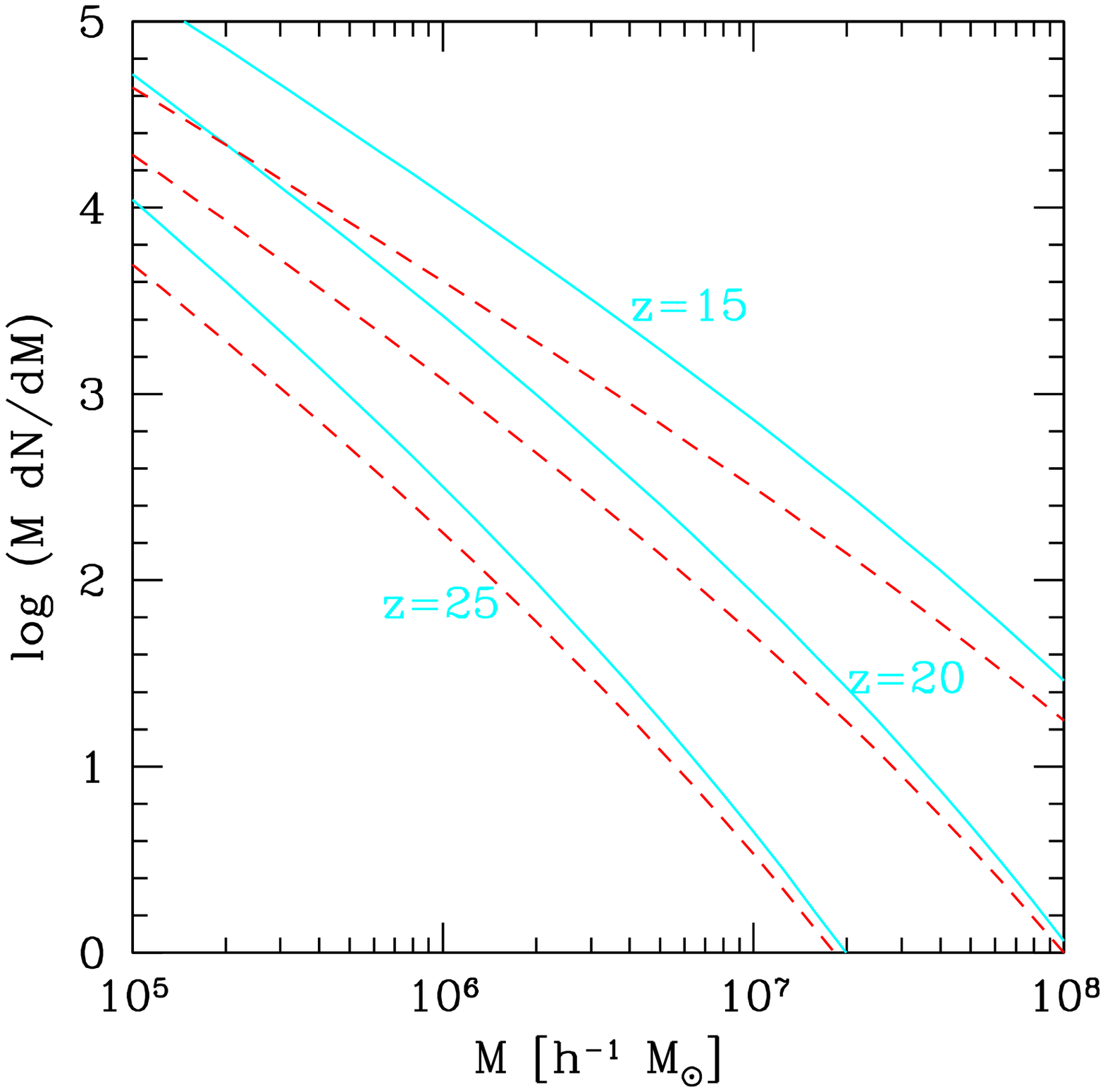,width=3.1in}}
\vspace{-0.3cm}
\caption{\footnotesize Mass function of minihalos formed at 
$z=15, 20, 25$ which, by the later time $z_0$, will have merged into 
a more massive halo of total mass $M_0$. {\it Solid curves:} $z_0=0.8$, 
$M_0=10^{12}\,h^{-1}\,\msun$ (`Milky Way' halo). {\it Dashed curves:} 
$z_0=3.5$, $M_0=2\times 10^{11}\,h^{-1}\,\msun$ (older `bulge'). 
} 
\label{fig1}
\vspace{+0.5cm}
\end{figurehere}

Consider now the fate of a population of $\sim$ 5,000 MBHs in 
the bulge of a `Milky Way' galaxy. We assume that the bulge mass -- $M_{\rm 
bulge}=2\times 10^{10}\,\msun$ within a radius $r_{\rm bulge}=1.5$ kpc -- 
is dominated by stars distributed with a singular isothermal profile of 
circular speed $V_c$, $\rho_\star(r)=V_c^2/(4 \pi G r^2)$,\footnote{This is 
similar to the distribution of stars observed in the inner few hundred pc of 
the Galactic center, $\rho_\star\propto r^{-1.8}$ (e.g. Genzel, Hollenbach, \& 
Townes 1994).}\,
truncated at $r_{\rm bulge}$. In our model this 
density profile is composed of stars, stellar remnants (including low--mass
black holes), and primordial MBHs.
Because of their large mass, MBHs `relax' much faster than ordinary stars 
and mass segregation occurs owing to dynamical friction.
A hole at radius $r=10\,r_{10}\,$pc with velocity $V_c=250\,V_{250}\,\kms$ 
will sink towards the center on the timescale
\beq
t_{\rm df}={1.17\over \ln \Lambda}\,{r^2V_c\over Gm_\bullet}={10^{10.64}\,{\rm yr}\over \ln\Lambda}\,r_{10}^2\,V_{250}\,\left({150\,\msun\over 
m_\bullet}\right) \label{tdf}
\eeq
(e.g. Binney \& Tremaine 1987; Begelman, Blandford, \& Rees 1980).
Equation (\ref{tdf}) with a Coulomb logarithm
of $\ln\Lambda=14$ predicts that any MBH whose original orbit lays within 
$r_{\rm df}=20\,(m_\bullet/150\,\msun)^{1/2}$ pc 
will spiral into the center in less than a Hubble time. A central cluster of 
$N_{\rm mbh}=5,000\,(r_{\rm df}/r_{\rm bulge})\gta 50\,(m_\bullet/150\,\msun)^{1/2}$
MBHs will form with a rate of infall which decreases as $t^{-1/2}$ and is currently
of order {\it two holes per Gyr per galaxy}. Here the precise numbers depend on the
stellar density profile and velocity distribution in the inner regions, and on 
the hole mass. Any increase in dynamical friction due to extra material 
(gas or dark matter) associated with the hole, either at the present--epoch
or -- perhaps more likely -- during the assembly of the halo from smaller
units, would enhance the central concentration and the number of MBHs near the 
nucleus.

The presence of a small central cluster of pregalactic MBHs would have 
several interesting
observational consequences, which will be discussed in details in a subsequent 
paper. Here we just want to point out that a close encounter 
may result in the tidal capture of an ordinary main--sequence star of mass 
$m_\star$ and radius $R_\star$ by a MBH, and the formation of a close 
binary system. Tidal capture occurs in galaxy cores whenever a star has a 
pericenter passage at $r_p=xR_\star$ very close to the tidal--breakup radius 
$r_t=(m_\bullet/m_\star)^{1/3}\,R_\star$. The tidal capture cross--section can be 
computed from the requirement that the energy stored in the tides, $\approx 
Gm_\bullet^2/(R_\star x^6)$ is larger than the relative energy of the pair 
at infinity (equal to $m_\star v_{\rm rel}^2/2$ for $m_\star\ll m_\bullet$), which 
is only a small fraction of the kinetic energy 
available at closest approach. This condition yields $x\lta (m_\bullet 
v_\star/m_\star v_{\rm rel})^{1/3}$, where $v_\star=(2Gm_\star/
R_\star)^{1/2}$ is the escape velocity at stellar surfacer. The cross--section 
leading to an encounter closer than $x$ is 
$\Sigma=\pi R_\star^2\,(v_\star/v_{\rm rel})^{7/3}\,(m_\bullet/m_\star)^{4/3}$ 
(Fabian, Pringle, \& Rees 1975), in good agreement with that derived 
by a detailed computation of the amount of 
energy deposited into oscillatory modes during a close encounter with a $n=1.5$
polytrope (Kim \& Lee 1999). For an isothermal stellar
distribution with one--dimensional velocity dispersion $\sigma$, the tidal capture 
rate at the galaxy center is then 
\beq
\Gamma={\rho_\star(0)\,N_{\rm mbh}\over m_\star (4\pi\sigma^2)^{3/2}}\int_0^\infty 
v_{\rm rel}\Sigma(v_{\rm rel}) \exp(-{v_{\rm rel}^2\over 4\sigma^2})d^3v_{\rm rel}.
\eeq
With $\sigma=V_c/\sqrt{2}=180\,\kms$, $m_\star=0.5\,\msun$, $R_\star=0.5\,\rsun$,
and $\rho_\star(0)=10^6\,\msun\,$pc$^{-3}$, one derives a rate of 
{\it one capture every $10^7\,$ yr}. The time a MBH would have to wait 
before tidally capturing a star is, assuming a typical relative velocity at infinity
$v_{\rm rel}^2=6\sigma^2$, longer than the dynamical friction timescale by a factor 
$\ln\Lambda\,(m_\star/m_\bullet)^{1/3}\,(v_\star/\sigma)^{5/3}$, 
and shorter than the capture timescale of a stellar hole by a factor 
$m_\bullet^{4/3}\gta 800$.   
Because of the high velocity dispersion, the binaries formed in galaxy
centers will be much closer than those formed in the cores of globular clusters.
Indeed, close encounters leading to the tidal disruption of the main--sequence
star (initial pericenter separation $r_p\lta r_t$ or $v_{\rm rel}\gta 
v_\star\approx 600\,\kms$) will be very common. Even when $r_p\gta r_t$,
tidal effects during each subsequent periastron passage will tend to 
circularize the orbit, and the high tidal luminosity will cause the star
to expand and overflow its Roche lobe.
Thus we can reasonably expect that tidally--captured stars will undergo 
some form of rapid mass loss (disruption) or mass transfer. If the debris are
accreted with high radiative efficiency by the MBH, then this may naturally 
lead to an off--center (`non--AGN') source with a peak X--ray luminosity close to
the Eddington limit, $2\times 10^{40}\,(m_\bullet/150\,\msun)\,\ergss$, shining
over a timescale of order $1.4\times 10^5\,(m_\bullet/150\,\msun)^{-1}\,$ yr.
Non--nuclear compact X--ray sources with such high luminosities have been 
observed, e.g., by {\it ROSAT} (Roberts \& Warwick 2000).

Finally, we draw attention to the fact that MBHs sinking towards the center
may be captured by the supermassive black hole (SMBH) in the nucleus and removed 
from the cluster as they diffuse into low--angular momentum `loss--cone' orbits.
They would then be swallowed either by direct scattering into the SMBH or by gradual
shrinking of their orbits due to gravitational wave emission. 
The case for stellar holes has been discussed by Sigurdsson \& Rees (1997)
and Miralda--Escud\'e \& Gould (2000). The rate of orbit decay through 
gravitational radiation is proportional to $m_\bullet$, and so is the 
characteristic strain amplitude of a wave emitted by a MBH in a given orbit around 
a SMBH of given mass (for $m_\bullet\ll M_{\rm smbh}$). The capture of a 
$150\,\msun$ hole
by a central black hole with $M_{\rm smbh}\approx 10^6\,\msun$ would be  
detectable by the planned {\it Laser Interferometer Space Antenna} ({\it LISA})
out to high redshifts.

\acknowledgments
\ni We have benefited from informative discussions with A. Ferrara, 
X. Hernandez, M. Salvati, and S. Woosley. 
Support for this work was provided by NASA through ATP grant NAG5--4236 
and a B. Rossi Visiting Fellowship at the Observatory of Arcetri (P.M.),
by the Royal Society (M.J.R.), and by the EC RTN network ``The Physics of the 
Intergalactic Medium''.

\end{document}